\documentclass[letterpaper,twocolumn,english,aps]{revtex4}
\usepackage[T1]{fontenc}
\usepackage[latin9]{inputenc}
\usepackage{array}
\usepackage{multirow}
\usepackage{amsmath}
\usepackage{amssymb}
\usepackage{graphicx}

\makeatletter

\pdfpageheight\paperheight
\pdfpagewidth\paperwidth

\providecommand{\tabularnewline}{\\}

\@ifundefined{textcolor}{}
{%
 \definecolor{BLACK}{gray}{0}
 \definecolor{WHITE}{gray}{1}
 \definecolor{RED}{rgb}{1,0,0}
 \definecolor{GREEN}{rgb}{0,1,0}
 \definecolor{BLUE}{rgb}{0,0,1}
 \definecolor{CYAN}{cmyk}{1,0,0,0}
 \definecolor{MAGENTA}{cmyk}{0,1,0,0}
 \definecolor{YELLOW}{cmyk}{0,0,1,0}
 }

\makeatother

\usepackage{babel}
\begin{document}

\title{Optimizing the second hyperpolarizability with minimally-parametrized
potentials}

\author{C. J. Burke}

\affiliation{Department of Physics and Astronomy, Center for Nanoscopic Physics,
Tufts University, 4 Colby Street, Medford, MA. 02155}

\author{J. Lesnefsky}

\affiliation{Department of Physics, University of Illinois at Chicago, 845 W.
Taylor St., Chicago, IL 60607-7059}

\author{R. G. Petschek}

\affiliation{Department of Physics, Case Western Reserve University, 10900 Euclid
Avenue, Cleveland, Ohio, USA 44106}

\author{T. J. Atherton}

\affiliation{Department of Physics and Astronomy, Center for Nanoscopic Physics,
Tufts University, 4 Colby Street, Medford, MA. 02155}
\begin{abstract}
The dimensionless zero-frequency intrinsic second hyperpolarizability
$\gamma_{int}=\gamma/4E_{10}^{-5}m^{-2}(e\hbar)^{4}$ was optimized
for a single electron in a 1D well by adjusting the shape of the potential.
Optimized potentials were found to have hyperpolarizabilities in the
range $-0.15\lessapprox\gamma_{int}\lessapprox0.60$; potentials optimizing
gamma were arbitrarily close to the lower bound and were within $\sim0.5\%$
of the upper bound. All optimal potentials posses parity symmetry.
Analysis of the Hessian of $\gamma_{int}$ around the maximum reveals
that effectively only a single parameter, one of those chosen in the
piecewise linear representation adopted, is important to obtaining
an extremum. Prospects for designing new chromophores based on the
design principle here elucidated are discussed.
\end{abstract}


\maketitle

\section{Introduction}

Developing materials with high electronic nonlinear susceptibilities
is of fundamental importance for a wide variety of applications such
as optical solitons, phase conjugate mirrors and optical self-modulation
\cite{boydbook,shenbook}. These susceptibilities are defined by considering
a material in the presence of an electric field $E$ and expanding
the induced polarization in a Maclaurin series,
\begin{equation}
P=\alpha E+\beta EE+\gamma EEE+O(E^{4}),\label{eq:gammadefinition}
\end{equation}
where the susceptibilities $\alpha$, $\beta$ and $\gamma$ are generally
frequency-dependent tensor quantities. Here, we consider the zero-frequency
limit of these quantities in the non-resonant regime i..e. where all
frequencies are much less than any resonant frequency of the electrons
and motion of the nuclei is neglected. A remarkable result due to
Kuzyk \cite{Kuzyk:2000p2905} is that quantum mechanics requires that
the first and second hyperpolarizabilities $\beta$ and $\gamma$
are bounded: specifically, $\gamma$ obeys the inequality, 

\begin{equation}
-\left(\frac{e\hbar}{\sqrt{m}}\right)^{4}\frac{N^{2}}{E_{10}^{5}}\le\gamma\le4\left(\frac{e\hbar}{\sqrt{m}}\right)^{4}\frac{N^{2}}{E_{10}^{5}}\equiv\gamma_{0}^{\text{max}},\label{eq:gamma}
\end{equation}
where $N$ is the number of electrons, $E_{10}$ is the energy difference
between the ground and first excited states and $m$ is the electron
mass. It is natural to define the intrinsic hyperpolarizability as
a figure of merit to characterize the proximity of a given system
to this limit,
\begin{equation}
\gamma_{int}=\gamma/\gamma_{0}^{\text{max}}\label{eq:gammaint}
\end{equation}
and to ask: how to create materials that achieve optimal $\gamma_{int}$?
The discovery of the bounds (\ref{eq:gamma}) has motivated a number
of experimental studies that have demonstrated that carefully tuning
the electronic states and geometry of chromophores can lead to higher
second hyperpolarizabilities\cite{Fujiwara:2008p3308,Hales:2010p3311,Luu:2005p3300,May:2007p3301}.
Generic design principles motivated by fundamental theory would therefore
be desirable. Unfortunately, the procedure used to derive the bounds
(\ref{eq:gamma}) cannot directly provide these; they were obtained
by optimizing $\gamma$ for a three-level ansatz with respect to the
dipole matrix elements and energy level spacings $E=E_{10}/E_{20}$
and \emph{not} by constructing an explicit potential. Indeed, the
assumptions behind the derivation have been questioned\cite{Champagne:2005p3194,Champagne:2006p3193}
and the limits need not be achievable with a local potential; it has
been speculated recently these may require exotic Hamiltonians\cite{Watkins:2012p3313}. 

Subsequent work, following approaches developed in earlier studies
of $\beta$\cite{Tripathy:2004p2912,Zhou:2007p2904}, has attempted
to address this in two ways: First, by identifying universal features
of Hamiltonians near the fundamental limit by a Monte Carlo search\cite{Shafei:2010p3304}
and, secondly, by numerically optimizing $\gamma_{int}$ with respect
to the shape of a local potential\cite{Watkins:2012p3313}. This latter
work found potentials which have second hyperpolarizabilities in the
range $-0.15\le\gamma_{int}\le0.60$, which represents an \emph{apparent}
bound that is more restrictive than the bound of (\ref{eq:gamma}).
Moreover it was demonstrated that the optimized potentials spectra
and dipole moments were broadly consistent with those identified in
the earlier Monte Carlo study. 

While these strategies provide useful goals for chemists attempting
to design new nonlinear chromophores, they do not provide insight
into which features of the potential are necessary to optimize $\gamma_{int}$,
or how many free parameters should be necessary to achieve optimal
or near optimal $\gamma$. In a previous paper\cite{Atherton:2012p3248},
we developed a technique to examine the analogous question for $\beta$:
by optimizing potentials described by increasing numbers of free parameters
and examining the eigenvalues of the Hessian matrix at each maximum,
we identified the combinations of parameters most important to the
optimization. The analysis revealed that effectively only two parameters
were necessary to maximize $\beta$, and hence that a surprisingly
broad range of potentials with high $\beta$ exists around each maximum. 

In this work, we apply the same technique to the problem of optimizing
$\gamma$. At first sight, the problem appears to be more difficult
than that for $\beta$ since the expression for $\gamma$ is much
more complicated and the bounds for positive and negative $\gamma$
are different. Remarkably, however, we will show that effectively
only \emph{one} parameter is necessary to optimize $\gamma$ in either
direction and, moreover, that in each case it is one of the parameters
utilized in our representation of the potential. In this sense, we
are able to suggest much more clearly a possible design strategy for
materials with high $\gamma$ than for $\beta$ within the limitations
of the model. At least within our representation of the potentials,
we find that the potential that maximizes $\gamma$ is rapidly varying,
while for negative $\gamma$, as for $\beta$, quite generic, slowly
varying potentials are adequate. The paper is organized as follows:
in section II the calculations performed are described; the results
are presented and discussed in section III; conclusions are drawn
in section IV.

\section{Model\label{sec:Model}}

It is first necessary to generalize the method described our previous
paper on optimizing the intrinsic first hyperpolarizability $\beta_{int}$
\cite{Atherton:2012p3248}: in the present work, $\gamma_{int}$ is
to be optimized by adjusting the shape of a one-dimensional piecewise-linear
potential. Such a potential with $N+1$ segments may be represented,
\begin{equation}
V(x)=\begin{cases}
A_{0}x+B_{0} & x<x_{0}\\
A_{n}x+B_{n} & x_{n-1}<x<x_{n},\ n\in\{1,...N-1\}\\
A_{N}x+B_{N} & x>x_{N-1},
\end{cases}\label{eq:arbpotential}
\end{equation}
with the positions $x_{n}$ and slopes $A_{n}$ as the adjustable
parameters and where the $B_{n}$ are chosen to enforce continuity.
Because $\gamma_{int}$ is invariant under trivial translations and
rescalings of the potential, some of these parameters were fixed $x_{0}=0$,
$B_{0}=B_{1}=0$, and $A_{1}=\pm1$. These choices, together with
a change of origin and rescaling allow for any potential. Thus maximizing
with these constraints allows faster optimization. Furthermore, the
left- and right-most slopes are required to be negative and positive,
respectively, ensuring only bound electron states. Finally, for technical
reasons, having to do with the asymptotic behavior of the Airy functions
introduced in eqn \ref{eq:airyfunctions} below, it is difficult to
allow the sign of any slope to change during an optimization. In consequence,
we have chosen to restrict $\mid A_{i}\mid>.005$, and, as appropriate
to do separate optimizations for each interesting sign of each slope.

A second representation for the potential was also considered where
parity symmetry was specifically enforced. This was motivated by previous
work\cite{Kuzyk:1990p3305} which identifies parity as important for
optimizing $\gamma_{int}$, particularly for the lower bound. The
potentials with enforced $\mathcal{P}$ symmetry were constructed
on the half line $x\ge0$ with $N$ segments,
\begin{equation}
V(x)=\begin{cases}
A_{n}x+B_{n} & x_{n-1}<x<x_{n},\ n\in\{1,...N-1\}\\
A_{N}x+B_{N} & x>x_{N-1}
\end{cases}\label{eq:psymmpotential}
\end{equation}
with $x_{0}=0$ and requiring $V(-x)=V(x)$. Again $x_{n}$ and $A_{n}$
are adjustable parameters and $x_{0}=0$, $B_{0}=B_{1}=0$ are fixed.
The parameter $A_{1}$ was set to either $-1$ or $+1$ to study the
consequences of both cases.

For such a potential with a uniform applied electric field of strength
$\epsilon$, the wavefunction obeys the Schrodinger equation in each
segment, 
\begin{equation}
\left[-\frac{1}{2}\frac{\text{d}^{2}}{\text{d}x^{2}}+(A_{n}+\epsilon)x+B_{n}\right]\psi_{n}=E\psi_{n},\label{eq:schrodinger}
\end{equation}
in units such that $e=1$, $\hbar=1$, and $m_{e}=1$. The solution
in each segment is written in terms of the well-known Airy functions,
\begin{eqnarray}
\psi_{n}(x) & = & C_{n}\text{Ai}\left[\frac{\sqrt[3]{2}(B_{n}-E+x(A_{n}+\epsilon))}{(A_{n}+\epsilon)^{2/3}}\right]+\nonumber \\
 &  & +D_{n}\text{Bi}\left[\frac{\sqrt[3]{2}(B_{n}-E+x(A_{n}+\epsilon))}{(A_{n}+\epsilon)^{2/3}}\right].\label{eq:airyfunctions}
\end{eqnarray}
To solve for the coefficients $C_{n}$ and $D_{n}$ the usual boundary
conditions are imposed, i.e. that the wavefunction $\psi(x)$ and
its derivative $\psi'(x)$ are continuous at the boundary between
segments. Additionally, in the end segments, the wavefunction must
vanish as $x$ goes to $\pm\infty$ fixing $D_{N}=0$. There are a
total of $2N$ linear equations in the coefficients for the arbitrary
case and $4N-2$ equations and coefficients for the $\mathcal{P}$-symmetric
case, which can be written in matrix form
\begin{equation}
W\cdot u=0\label{eq:Wueq0}
\end{equation}
where $u$ is a vector comprised of the $C_{n}$ and $D_{n}$ coefficients
and $W$ is a matrix which depends on $E$, $\epsilon$ and the parameters
$A_{n}$ and $x_{n}$.

The allowed energy levels are found by numerically finding the roots
of 
\begin{equation}
\det W=0
\end{equation}
with $\epsilon=0$. It is readily possible, as previously done for
$\beta$, to obtain from (\ref{eq:Wueq0}) an expression for the second
hyperpolarizability,
\begin{equation}
\gamma\equiv\frac{1}{6}\left.\frac{\mathrm{d}^{4}E_{0}}{\mathrm{d}\epsilon^{4}}\right|_{\epsilon=0};
\end{equation}
this is achieved by repeatedly differentiating the matrix $W$ using
the Jacobi formula,
\begin{equation}
\frac{\mathrm{d}}{\mathrm{d}\epsilon}\det W=\mathrm{Tr}\left(\mathrm{adj}W\cdot\frac{\mathrm{d}W}{\mathrm{d}\epsilon}\right),
\end{equation}
where $\mathrm{adj}W$ is the adjugate of $W$ (since $W$ is singular),
and applying the chain rule
\begin{equation}
\frac{\mathrm{d}W}{\mathrm{d}\epsilon}=\frac{\partial W}{\partial\epsilon}+\frac{\partial W}{\partial E}\frac{\mathrm{d}E}{\mathrm{d}\epsilon}.
\end{equation}
Having performed similar calculations to those in \cite{Atherton:2012p3248},
we arrive at the expression

\begin{widetext}
\begin{equation}
\frac{\mathrm{d}^{4}E}{\mathrm{d}\epsilon^{4}}=-\frac{\mathrm{Tr}\left[\left(\frac{\mathrm{d}^{3}}{\mathrm{d}\epsilon^{3}}\mathrm{adj}W\right)\cdot\frac{\mathrm{d}W}{\mathrm{d}\epsilon}+3\left(\frac{\mathrm{d}^{2}}{\mathrm{d}\epsilon^{2}}\mathrm{adj}W\right)\cdot\frac{\mathrm{d}^{2}W}{\mathrm{d}\epsilon^{2}}+3\left(\frac{\mathrm{d}}{\mathrm{d}\epsilon}\mathrm{adj}W\right)\cdot\frac{\mathrm{d}^{3}W}{\mathrm{d}\epsilon^{3}}+\mathrm{adj}W\cdot W'''\right]}{\mathrm{Tr}\left(\mathrm{adj}W\cdot\frac{\partial W}{\partial E}\right)},
\end{equation}
\end{widetext}where 
\begin{equation}
W'''=\frac{d^{4}W}{d\epsilon^{4}}-\frac{\partial W}{\partial E}\frac{d^{4}E}{d\epsilon^{4}}.
\end{equation}
From this, $\gamma$ is readily obtained and the intrinsic second
hyperpolarizability $\gamma_{int}=\gamma/\gamma_{0}^{\textrm{max}}$
calculated for a given set of parameters.

The quantity $\gamma_{int}$ was optimized numerically for both arbitrary
and $\mathcal{P}$-symmetric potentials with varying numbers of segments
using the \texttt{FindMaximum} function of \emph{Mathematica} 8, an
implementation of the Interior Point method for constrained optimization.
Both maxima and minima of $\gamma_{int}$ were obtained from a large
number of randomly generated starting points, and also manually chosen
starting points with large values for $\gamma_{int}$. Once an optimum
$\gamma_{int}$ was found, the extent to which each of the parameters
was important to the extremum was characterized by calculating the
Hessian matrix,
\[
H_{ij}=\frac{\partial^{2}}{\partial P_{i}\partial P_{j}}\gamma_{int},
\]
where the $P_{i}$ are the parameters, and calculating its eigenvalues
and eigenvectors. Since the Hessian matrix characterizes the local
curvature of the objective function in the parameter space around
the extremum, these quantities give the magnitudes and directions
of the principal curvatures. As stressed in previous work\cite{Atherton:2012p3248}
these curvatures implicitly depend on a measure implied by this equation
that is peculiar to our numerical parameterization of the problem.
More physically relevant measures can also be used to calculate eigenvalues.
While these make quantitative changes in the eigenvalues and vectors,
they do not make qualitative changes, and we give results for this
{}``numerically natural'' measure below.

\section{Results And Discussion}

\begin{figure}
\centering{}\includegraphics{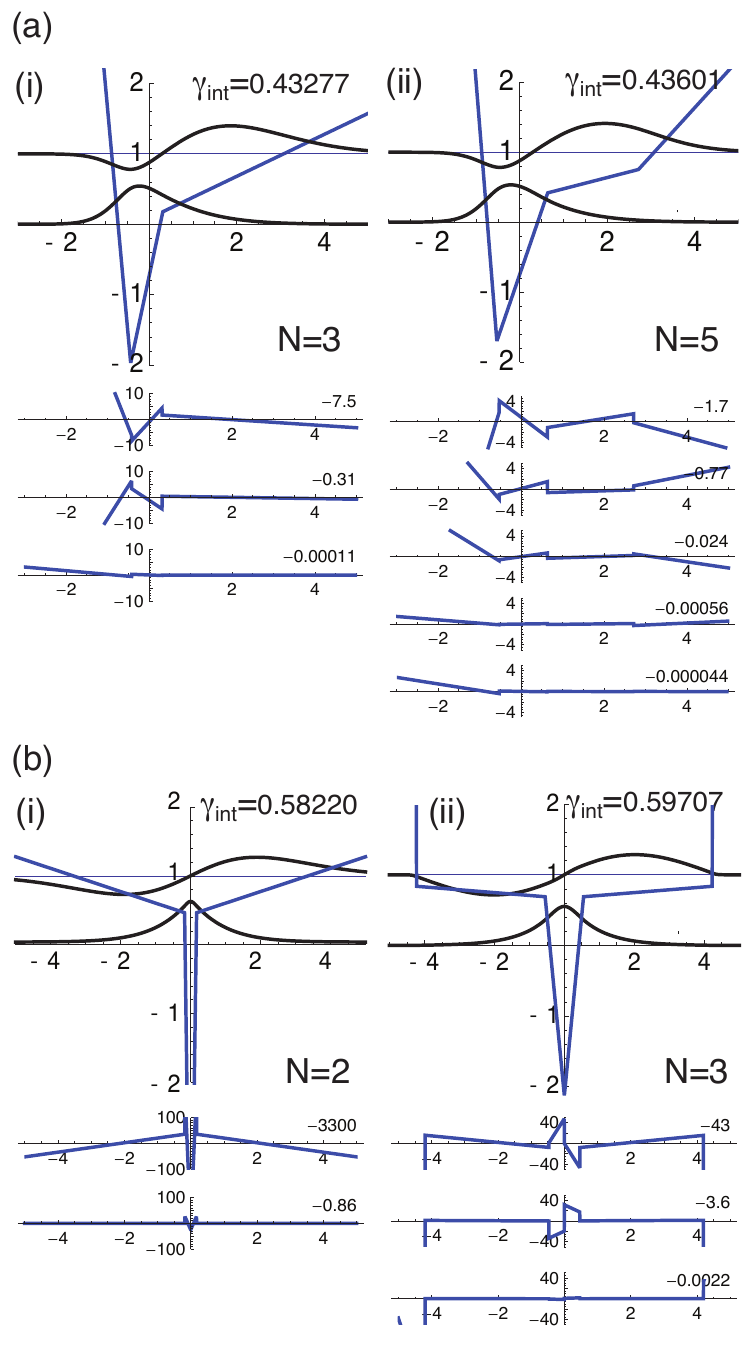}\caption{\label{fig:Results-for-all}Optimized potentials for the \emph{upper}
bound of $\gamma_{int}$ with (a) no enforced symmetry and (b) $\mathcal{P}$-symmetry.
The energies of the ground and first excited state are indicated by
horizontal lines; the corresponding wavefunctions are also displayed.
The plots have been rescaled to facilitate comparison by ensuring
$E_{1}-E_{0}=1$ and $\left\langle x_{0}\right\rangle =0$ while preserving
$\gamma_{int}$. }
\end{figure}

\begin{figure}
\centering{}\includegraphics{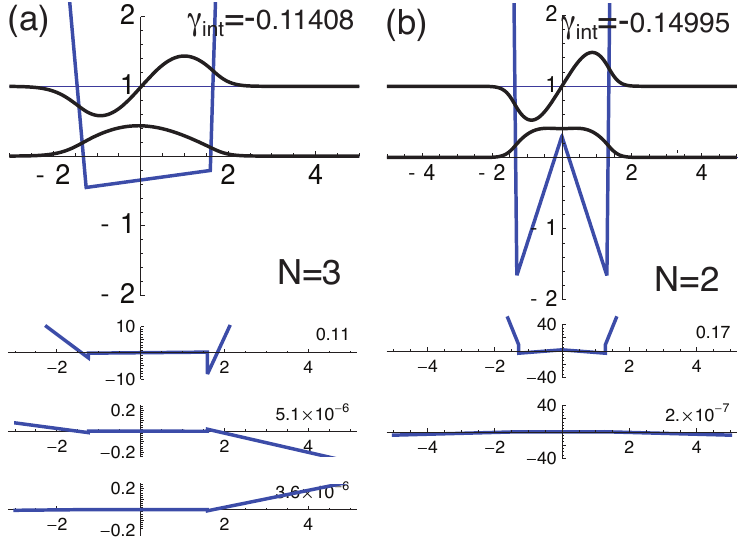}\caption{\label{fig:ResultsPsymmetry}Optimized potentials for the \emph{lower}
bound of $\gamma_{int}$ with (a) no enforced symmetry and (b) $\mathcal{P}$-symmetry. }
\end{figure}

\begin{figure}
\centering{}\includegraphics{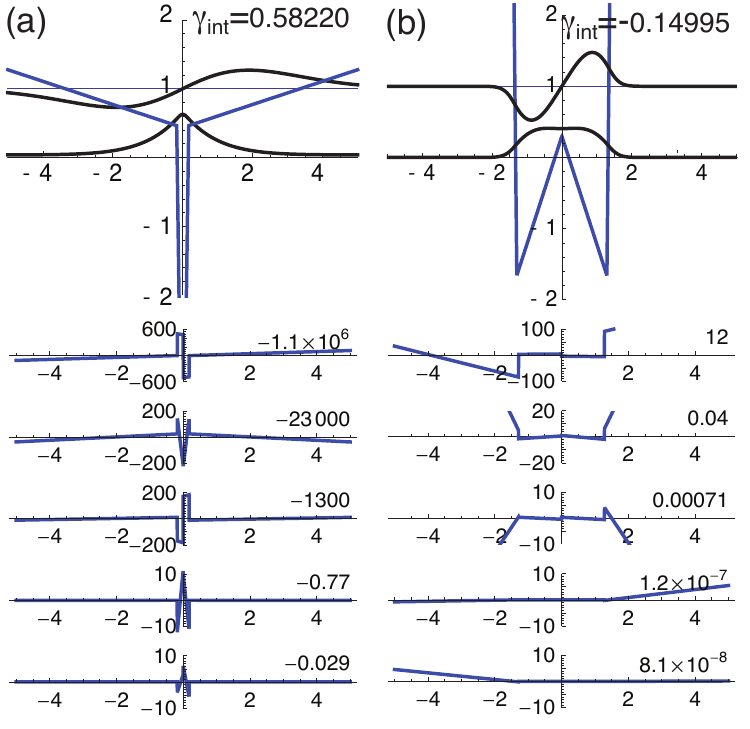}\caption{\label{fig:ResultsExpandedEigenvectorspace}Five parameter potentials
with no enforced symmetry optimized using two-parameter $\mathcal{P}$-symmetric
optima as the starting point. (a) Upper and (b) lower bounds.}
\end{figure}

\begin{table*}
\begin{tabular}{ccccccccc}
\hline 
Bound & Description & Ref. & $\gamma_{int}$ & $A_{0}$ & $A_{2}$ & $A_{3}$ & $x_{1}$ & $x_{2}$\tabularnewline
\hline 
\hline 
\multirow{4}{*}{Upper} & 3 param. arb. & (a)(i) & $0.43277$ & $2.33932$ & $0.10189$ & \textemdash{} & $1.04960$ & \textemdash{}\tabularnewline
 & 5 param. arb. & (a)(ii) & $0.43602$ & $4.28494$ & $0.08829$ & $0.37688$ & $1.42131$ & $3.95728$\tabularnewline
 & 2 param. $\mathcal{P}$ & (b)(i) & $0.58220$ & \textemdash{} & $0.00550$ & \textemdash{} & $0.57426$ & \textemdash{}\tabularnewline
 & 3 param. $\mathcal{P}$ & (b)(ii) & $0.59707$ & \textemdash{} & $0.00500$ & \textemdash{} & $0.88533$ & $7.95755$\tabularnewline
\cline{2-9} 
\multirow{2}{*}{Lower} & 3 param. arb. & (a)(iii) & $-0.11409$ & $82.471$ & $237.57$ & \textemdash{} & $1.25227$ & \textemdash{}\tabularnewline
 & 2 param. $\mathcal{P}$ & (b)(iii) & $-0.14996$ & \textemdash{} & $31.6157$ & \textemdash{} & $1.48005$ & \textemdash{}\tabularnewline
\hline 
\end{tabular}

\caption{\label{tab:ParameterValues}List of parameters of optimized potentials}
\end{table*}

Our optimized potentials, together with the ground and first excited
state wavefunctions, are displayed in Fig. \ref{fig:Results-for-all}
for both arbitrary (eq. \ref{eq:arbpotential}) and $\mathcal{P}$-symmetric
(eq. \ref{eq:psymmpotential}) parametrizations. The associated parameter
values are listed in Table \ref{tab:ParameterValues}. The potentials
and wavefunctions are displayed on a transformed position and energy
scale,
\begin{eqnarray}
\bar{x} & = & (x-\left\langle x\right\rangle )/(E_{1}-E_{0})^{1/2}\nonumber \\
\bar{V}(\bar{x},\{P\}) & = & (V(\bar{x},\{P\})-E_{0})/(E_{1}-E_{0})\label{eq:scaling}
\end{eqnarray}
such that the ground state energy is $E_{0}=0$, the difference between
the ground and first excited state energy is $E_{1}-E_{0}=1$ and
the position expectation value for the ground state is $\left\langle x_{0}\right\rangle =0$.
This rescaling does not change $\gamma_{int}$ and permits convenient
comparison of the results of each optimization. To identify the relative
importance of each of the parameters to the optimization, the results
of the eigenanalysis of the Hessian matrix are also displayed for
selected potentials in Fig. \ref{fig:Results-for-all}; the $j$-th
eigenvalue of the Hessian, $h^{j}$, is listed alongside a plot of
the variation in the potential $\Delta V^{j}(x)$ in the direction
of the associated eigenvector, 
\begin{equation}
\Delta V^{j}(x)=\left.\frac{\partial\bar{V}(\bar{x},\{P_{i}+\alpha v_{i}^{j}\})}{\partial\alpha}\right|_{\alpha=0},\label{eq:deltavj}
\end{equation}
where $v_{i}^{j}$ is the $i^{th}$ component of the $j$-th eigenvector.
Note that the values of $V$ and $x$ in the right hand side of (\ref{eq:deltavj})
are renormalized as a function of $\alpha$ using (\ref{eq:scaling})
so that the variations presented automatically preserve the properties
$\left\langle x_{0}\right\rangle =0$ and $E_{1}-E_{0}=1$. 

The first set of results displayed in Fig. \ref{fig:Results-for-all}(a)
are optimized potentials with no enforced symmetry and specified by
3 or 5 free parameters. The optimized $\gamma_{int}$ for both the
lower and upper bounds of $\gamma_{int}$ potentials fall somewhat
below the apparent bounds observed in \cite{Shafei:2010p3304}. For
the upper bound, the best results assuming all slopes except the first
are positive are what appears to be a local maximum value of $\gamma_{int}\backsimeq0.43$
{[}Fig. \ref{fig:Results-for-all}(a)(ii){]}. A similar value was
also found in \cite{Watkins:2012p3313} for potentials with no constrained
symmetry, but the potentials found here do not closely resemble those
found in that work. The eigenvalues and eigenvectors displayed below
the potential in {[}Fig. \ref{fig:Results-for-all}(a)(ii){]} show
that only two of the eigenvalues are significant in magnitude and
are associated with the shape of the potential in the middle while
the small eigenvalues are associated with the outer slopes. These
results are reminiscent of those found for the first hyperpolarizability,
where $\beta_{int}$ was found to approach its maximum value for the
same class of potentials with a similarly small number of parameters
and analysis of the Hessian revealed that only effectively two parameters
were important to the maximization. For the lower bound, the 3 parameter
system {[}Fig. \ref{fig:ResultsPsymmetry}(a){]} converges on a shape
approaching a square well. The 5 parameter system also converges on
a $\mathcal{P}$-symmetric potential and, because of this, further
discussion of the lower bound is deferred to a subsequent paragraph. 

A possible explanation for the fact that the optimized positive $\gamma_{int}$
for arbitrary potentials in Fig. \ref{fig:Results-for-all} falls
short of the bounds established in \cite{Kuzyk:2000p2905,Shafei:2010p3304,Watkins:2012p3313}
is that the hyperpolarizability is sensitive only to some features
of the potential and that many local extrema exist. Rather than making
extensive runs starting from a variety of potentials, we chose to
use the more constrained subset of $\mathcal{P}$-symmetric potentials
(\ref{eq:psymmpotential}) and found that, indeed, even with only
2-3 parameters, much higher values of $\gamma_{int}$ could be obtained
as shown in Fig. \ref{fig:Results-for-all}(b). This is reminiscent
of the observation in \cite{Watkins:2012p3313} that these bounds
could only be reached if a $\mathcal{P}$-symmetric starting point
was used for the search; in this work, we not only enforce the symmetry
of the starting point but at all times in the optimization.

For the upper bound, we found a two-parameter potential close to the
apparent maximum but below the theoretical maximum {[}Fig. \ref{fig:Results-for-all}(b)(i){]}.
The shape is characterized by shallow outer slopes with a divot in
the center; the ground state wave function is localized to the divot
with the highly delocalized first excited state above the divot. It
was found to be necessary to constrain the slope $A_{2}>0.005$ since
the method of calculating $\gamma_{int}$ presented in section \ref{sec:Model}
fails for shallow slopes. To avoid the unphysical feature of delocalized
higher excited states, a distant wall was added to construct a three-parameter
potential {[}Fig. \ref{fig:Results-for-all}(b)(ii){]}. Such a change
is expected \emph{a priori} from previous work\cite{Wiggers:2007p2948}
to make no significant difference to $\gamma_{int}$ as it is far
from the region where $\psi_{0}$ and $\psi_{1}$ are large. For this
well-like potential, we performed a maximization, adjusting $A_{2}$,
$x_{1}$, and $x_{2}$ while fixing the outer walls to have a large
slope ($A_{3}=100$). The best potential found possesses $\gamma_{int}=0.59707$,
and has $A_{2}=0.005$ which is the shallowest slope allowed by the
constraint. The eigenvectors of the Hessian, calculated for the subspace
excluding $A_{2}$, are well aligned with the parameters: the most
significant eigenvector corresponds to $x_{1}$, the outer boundary
of the divot. The other eigenvector corresponds to $x_{2}$, the position
of the outer walls; this would be expected to have relatively little
influence on $\gamma_{int}$ since it controls a feature where the
ground state wavefunction is small.Fixing the outer slopes at $A_{3}=100$
as above, we also attempted to optimize $\gamma_{int}$ with $A_{2}$
constrained to be negative. It was found that $\gamma_{int}$ increased
as the slope approached zero, until a point was reached where the
calculations became numerically unstable due to the asymptotic properties
of the Airy functions. The highest value of $\gamma_{int}$ which
was found within a region of parameter space for which the calculation
was still stable was $\gamma_{int}=0.5915$, lower than the current
maximum. We then performed similar optimizations on potentials with
strict hard wall boundary conditions. The calculations for these potentials
\emph{were} numerically stable in all regions of parameter space which
were explored: For $A_{2}>0$, a maximum of $\gamma_{int}=0.5959$
was found. For $A_{2}<0$ a higher maximum of $\gamma_{int}=0.5968$
was found, though this is still lower than the current maximum of
$\gamma_{int}=0.5971$. Since a higher $\gamma_{int}$ is found in
potentials with $A_{2}<0$ than for potentials with positive $A_{2}$
in cases with strict hard wall boundary conditions, we speculate that
a value of $\gamma_{int}$ higher than the current maximum found might
be found for the finite $A_{3}$ case. Nonetheless, we do not expect
to see a significant improvement as the current maximum is already
within $\sim0.5\%$ of the maximum value found in previous studies.

For the lower bound, a potential with the best value of $\gamma_{int}=-0.1500$
was found using only two parameters {[}Fig. \ref{fig:ResultsPsymmetry}(b){]}.
This potential is characterized by steep outer walls and a {}``bump''
in the middle: the ground state and first excited state wavefunctions
cover the same spatial extent, but the bump causes the ground state
to become spread out and relatively flat. Eigenanalysis of the Hessian
of $\gamma_{int}$ about this solution shows that one eigenvalue is
significantly larger than the other, indicating that only one of the
parameters is physically relevant. Moreover, the eigenvectors of the
Hessian for this potential are aligned with the parameter space chosen
to represent the potential. The higher eigenvalue is associated with
an eigenvector along the $x_{1}$ direction, which determines the
position of the outer walls; the smaller eigenvalue is associated
with the parameter that controls the slope of the outer walls. Because
$\gamma_{int}$ is invariant under rescalings of the form (\ref{eq:scaling}),
a potential with identical $\gamma_{int}$ can be constructed for
a well of arbitrary width by tuning the slope of the bump. 

The two parameter $\mathcal{P}$-symmetric potentials identified as
the satisfying the apparent lower bound can be equivalently represented
by a five parameter arbitrary potential. Optimization of the five
parameter potential indeed finds this potential as the apparent global
maximum. There are therefore no directions in this new parameter space
which lead to a higher $\gamma_{int}$, despite relaxing the requirement
that $x_{-n}=-x_{n}$ and $A_{-n}=-A_{n}$. While the existence of
asymmetric potentials with more negative $\gamma_{int}$ cannot be
ruled out, our analysis confirms that a $\mathcal{P}$-symmetric potential
satisfies the apparent lower bound. We repeated this procedure for
the upper bound using the two parameter $\mathcal{P}$-symmetric potential
displayed in Fig. \ref{fig:ResultsPsymmetry}(b) as the starting point
for optimization. It was found that despite relaxing the symmetry
constraint no further improvement could be made so that the potential
of Fig. \ref{fig:ResultsPsymmetry}(b) is also a local optimum with
respect to the expanded parameter space. Analysis of the hessian was
also performed on both of these five parameter optima. The eigenvalues
and associated eigenvectors, shown in Fig. \ref{fig:ResultsExpandedEigenvectorspace},
reveal that for both upper and lower bounds the variation in the potential
associated with the largest eigenvalue is indeed asymmetric.

\begin{figure}
\centering{}\includegraphics[width=3.4in]{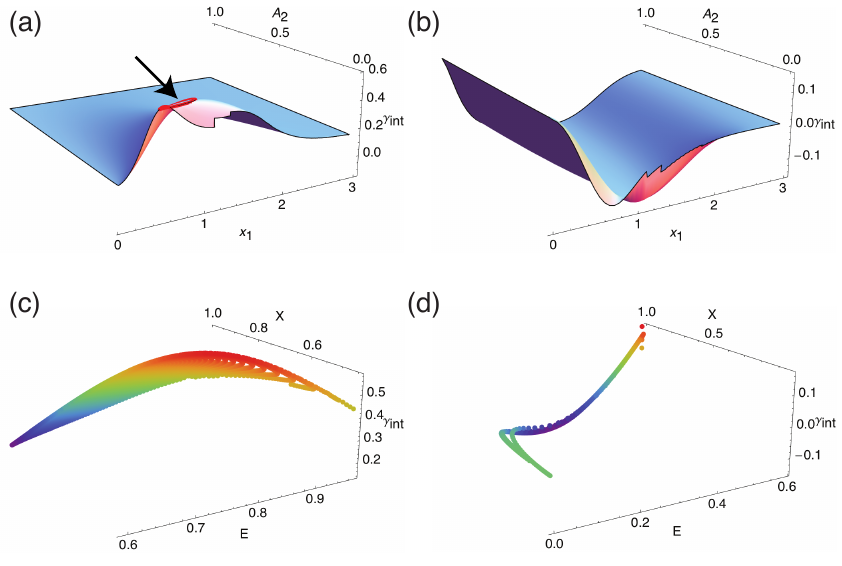}\caption{\label{fig:visparamspace}Visualization of the variation of $\gamma_{int}$
as a function of the parameters $x_{1}$ and $A_{2}$ of a two-parameter
$\mathcal{P}$-symmetric potential for the (a) upper bound and (b)
lower bound. The region of parameter space where $\gamma_{int}$ is
within 2\% of the maximum value is highlighted in (a) and indicated
by an arrow. $\gamma_{int}$ is also shown in the $(E,X)$ parameter
space for the (c) upper bound and (d) lower bound. }
\end{figure}

The fact that we are able to achieve optimized $\gamma_{int}$ within
1\% of previous limits with far fewer parameters than in other representations
of the potential\cite{Watkins:2012p3313}, and also fewer parameters
than required to numerically optimize $\beta_{int}$, is surprising
since the calculated expressions for $\gamma_{int}$ are far more
complicated than those for $\beta_{int}$. Since the dimensionality
of this parameter space is so small, it is instructive to visualize
it directly: For our two-parameter optimizations, the value of $\gamma_{int}$
is plotted over a portion of the parameter space {[}Fig. \ref{fig:visparamspace}(a)
and (b){]}. The plot for the lower bound merely illustrates the results
obtained from analysis of the Hessian, i.e. that the minimum is strongly
curved about the optimal value of $x_{1}$ but shallow with respect
to $A_{2}$. The plot for the upper bound is more interesting: the
region of parameter space for which $\gamma_{int}$ is within 2\%
of the maximum value obtained is highlighted showing a clear ridge. 

By calculating values of $X=x_{01}/x_{01}^{\text{max}}$ and $E=E_{01}/E_{02}$,
the natural parameters of the three-level ansatz\cite{Kuzyk:2000p2905},
as a function of $(A_{2},x_{1})$, we are able to display $\gamma_{int}$
re-parametrized in $(E,X)$ space {[}Fig. \ref{fig:visparamspace}(c)
and (d){]}. Here, $x_{01}^{max}=1/\sqrt{2E_{10}}$ in our units. Notice
that the entire region explored in the numerical parametrization $(A_{2},x_{1})$
collapses onto a narrow, elongated region in $(E,X)$ space for the
upper bound {[}Fig. \ref{fig:visparamspace}(c){]} and a complicated
curved line for the lower bound{[}Fig. \ref{fig:visparamspace}(d){]}.
These plots confirm the results of eigenanalysis of the Hessian: that
essentially only a single parameter (or 2 for the upper bound) characterizes
optimal $\gamma_{int}$. 

\begin{figure}
\centering{}\includegraphics[scale=0.66]{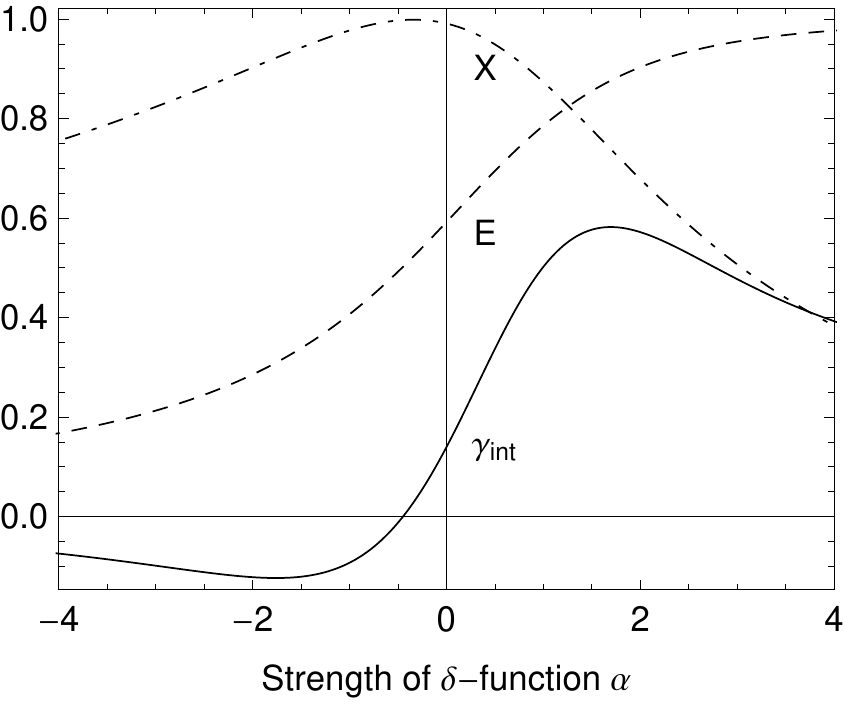}\caption{\label{fig:deltafunction}Plot of $\gamma_{int}$, $E=E_{10}/E_{20}$
and $X=x_{10}/x_{10}^{max}$ as a function of $\alpha$ for the for
the potential $V=|x|-\alpha\delta(x)$.}
\end{figure}

The results of the three-parameter optimization and the plot in fig.
(\ref{fig:visparamspace}) both suggests that the truly optimal $\mathcal{P}$-symmetric
potential for the upper bound has shallow outer slopes $A_{2}\to0$.
Such a potential can be transformed, using (\ref{eq:scaling}), to
a potential of equivalent $\gamma_{int}$ but where the outer slope
is unity and the central well is far narrower and sharper. Since the
central divot for the transformed potential resembles a Dirac delta
function, we studied the second hyperpolarizability of the family
of potentials
\begin{equation}
V(x)=\left|x\right|-\alpha\delta(x)\label{eq:deltapotential}
\end{equation}
where $\alpha$ is the single adjustable parameter. Values of $\gamma_{int}$
as a function of $\alpha$ are displayed in fig. \ref{fig:deltafunction}
and the maximum value is found to be $\gamma_{int}=0.58194$ which
occurs when $\alpha=1.69552$. Despite the simplicity of this one-parameter
potential, the result is only $3\%$ smaller than the best reported
so far and only fractionally smaller than those found with the two
and three parameter potentials above. 

We now turn to the question of whether the simple, but optimal or
nearly so, potentials that were identified above resemble either those
previously found \cite{Watkins:2012p3313} or possess the universal
features identified in \cite{Shafei:2010p3304}. Comparing our best
optimized potentials and wavefunctions to those in \cite{Watkins:2012p3313},
some qualitative similarities are apparent. For the upper bound, the
potentials in Watkins et al. are roughly symmetric near their lowest
point. They feature a central divot within a wider well where the
ground state wave function is localized within the central divot and
the first excited wave function is relatively delocalized compared
to the ground state. For the lower bound, the potentials feature a
steep well within which both the ground state and first excited state
are localized, and a central bump which causes the ground state to
be spread out within the well. These qualitative features are shared
by the potentials obtained in this work, but many extraneous details
are removed by the highly constrained, judiciously chosen representation.

In table \ref{tab:physical} we display values of $X$ and $E$, together
with the values of these parameters that extremize $\gamma_{int}$
for the three-level ansatz\cite{Kuzyk:2000p2905} and those possessed
by the best previously found potentials\cite{Watkins:2012p3313}.
Values are also shown for two elementary potentials, the triangle
well and infinite square well. The results for the upper bound are
quite consistent with those of Watkins \emph{et al.} if the breadth
in the range of these values identified by the Monte Carlo study\cite{Shafei:2010p3304}
is taken into account. The values of $E$ and $X$ are also displayed
in Fig. \ref{fig:deltafunction} as a function of $\alpha$, the strength
of the $\delta$ function, in the potential (\ref{eq:deltapotential});
$E$ is a monotonically increasing function of $\alpha$ while $X$
is a monotonically decreasing function. Crudely, these explain the
existence of a maximum $\gamma_{int}$ as representing the trade-off
between increasing the motion of the electron (associated with high
$X$) versus enhancing transitions to other states (associated with
low $E$). 

\begin{table}
\begin{tabular}{ccccc}
\hline 
Bound & Potential & $\gamma_{int}$ & $E$ & $X$\tabularnewline
\hline 
\multirow{6}{*}{Upper} & 3-level ansatz\cite{Kuzyk:2000p2905} & $1$ & $0$ & $0$\tabularnewline
 & Best from \cite{Watkins:2012p3313} & $0.5993$ & $0.5767$ & $0.5150$\tabularnewline
 & Arb. 5 param. & $0.43601$ & $0.5765$ & $0.6512$\tabularnewline
 & $\mathcal{P}$ 3 param. & $0.59707$ & $0.8665$ & $0.7468$\tabularnewline
 & $|x|-\alpha\delta$ & $0.58194$ & $0.6842$ & $0.7385$\tabularnewline
 & Triangle well $|x|$ & $0.1392$ & $0.5918$ & $0.9911$\tabularnewline
\cline{2-5} 
\multirow{5}{*}{Lower} & 3-level ansatz\cite{Kuzyk:2000p2905} & $-0.25$ & $0$ & $\pm1$\tabularnewline
 & Best from\cite{Watkins:2012p3313} & $-0.1500$ & $0.1493$ & $0.6658$\tabularnewline
 & $\mathcal{P}$ 2 param. & $-0.1500$ & $0.2855$ & $0.9416$\tabularnewline
 & $|x|-\alpha\delta$ & $-0.1236$ & $0.2438$ & $0.9222$\tabularnewline
 & Infinite square well & $-0.1262$ & $0.3750$ & $0.9801$\tabularnewline
\hline 
\end{tabular}

\caption{\label{tab:physical}Second hyperpolarizabilities and physical parameters
$X=x_{01}/x_{01}^{\text{max}}$ and $E=E_{01}/E_{02}$ for the optimized
potentials obtained in this work.}
\end{table}

\section{Conclusion}

We have optimized the intrinsic second hyperpolarizability $\gamma_{int}$
of a piecewise linear potential well with respect to parameters that
control the shape of the potential. We found solutions that lie within
the range $-0.15\le\gamma_{int}\lessapprox0.60$ in agreement with
the apparent bounds established in previous numerical optimizations\cite{Watkins:2012p3313};
these both fall short of the Kuzyk limits\cite{Kuzyk:2000p2905}.
By using two types of potential, one where all slopes were allowed
to vary and another with explicitly enforced symmetry, we demonstrated
that $\mathcal{P}$-symmetric potentials satisfy the apparent lower
bound for $\gamma_{int}$ and come within $\sim0.5\%$ of the apparent
upper bound. The parametrization used constrains the potential to
be smooth, preventing the occurrence of rapid oscillations which do
not affect $\gamma_{int}$ \cite{Wiggers:2007p2948}. Because of this
and the strong symmetry constraint, the optimal $\mathcal{P}$-symmetric
potentials found were characterized by only $2-3$ parameters. Of
these,\emph{ a posteriori} analysis of the Hessian revealed that effectively
only one or two, for the lower and upper bound respectively, were
important

These results are reminiscent of those obtained earlier for $\beta_{int}$\cite{Atherton:2012p3248},
yet the number of parameters required to optimize $\gamma_{int}$
appears to be smaller even though it is a more complex object, containing
more terms and involving higher derivatives. At least part of the
reason for this is that for $\gamma_{int}$ there exists a {}``compatible''
symmetry operator, $\mathcal{P}$, which can be used to constrain
the shape of the potentials; this was not possible for $\beta_{int}$
where $\mathcal{P}$-symmetric potentials automatically have $\beta=0$.
However, even though we have shown that $\mathcal{P}$-symmetric potentials
can have optimal or near-optimal second hyperpolarizabilities, it
is not clear whether the apparent upper bound $\gamma_{int}\sim0.6$
achieved by Watkins and Kuzyk can be achieved with a $\mathcal{P}$-symmetric
potential or whether a small amount of asymmetry is necessary. Moreover,
the reason why local potentials fall short of the Kuzyk bounds remains
opaque. 

The small parameter space allows us to propose a clear design paradigm
for new chromophores, within the limitations of the model one-electron
1D system studied. Neglected here, for example, are multi-electron
interactions, molecular ordering and inter-molecular electron hopping.
Nonetheless, the potentials obtained could be realized, for example,
by a centrosymmetric molecule with a central attractive or repulsive
group\textemdash{}for positive or negative $\gamma_{int}$ respectively.
The strength/electronegativity of the central group and the ratio
of the length of the central and peripheral groups can then be tuned
to give high $\gamma_{int}$. Unfortunately, most practical chromophores
are $\pi$ conjugated systems in which there are approximately as
many electrons as there are sites on the molecule, thus the consequences
for them from this single electron calculation are clearly very speculative.
Nevertheless, homologous sequences already studied for high $\gamma$
as a function of chain length, e.g. \cite{Luu:2005p3300} could likely
be enhanced by including such a central group with different electronegativity.
Any other means of achieving a potential well or inducing a significant
phase shift in wavefunctions passing through the center of the molecule
is also likely, even in multi-electron systems, to offer a route for
achieving larger $\gamma_{int}$. The present analysis provides other
important insights: first, that since the {}``true'' parameter space
for $\gamma_{int}$ is so small, only rough tuning of the molecular
design ought to be necessary. Secondly, this work again confirms that
there are a large set of modifications to optimized potentials, e.g.
rapid oscillations, that will not change $\gamma_{int}$ and need
not be considered in planning what molecules to synthesize. As has
been previously noted, materials with high $\gamma_{int}$ could also
be realized more directly in other ways, such as through composite
materials.

\end{document}